\documentclass[%
 reprint,
superscriptaddress,
 amsmath,amssymb,
 aps,
]{revtex4-1}
\usepackage{color}
\usepackage{braket}
\usepackage{bm}
\usepackage{enumitem}
\usepackage{todonotes}
\usepackage{graphicx}% Include figure files
\usepackage{dcolumn}% Align table columns on decimal point
\usepackage{subcaption}
\usepackage{bm}% bold math
\usepackage{hyperref}% add hypertext capabilities
\begin{document}

\preprint{APS/123-QED}

\title{Polaritonic frequency-comb generation and breather propagation in a negative-index metamaterial with a cold four-level atomic medium}% Force line breaks with \\

\author{Saeid Asgarnezhad-Zorgabad}
 \affiliation{Department of Physics, Sharif University of Technology, Tehran, 11365 11155, Iran}%Lines break automatically or can be forced with \\
 \affiliation{Institute for Quantum Science and Technology, University of Calgary, Calgary, Alberta T2N 1N4 Canada}
\author{Pierre Berini}
\affiliation{Department of Physics, University of Ottawa, 150 Louis-Pasteur, Ottawa, Ontario K1N 6N5, Canada }%
\affiliation{Centre for Research in Photonics, University of Ottawa, 25 Templeton St., Ottawa, Ontario K1N 6N5, Canada}%
\affiliation{School of Electrical Engineering and Computer Science, 700 King Edward St., Ottawa, Ontario, K1N 6N5 Canada}%
\author{Barry C. Sanders}
\affiliation{Institute for Quantum Science and Technology, University of Calgary, Calgary, Alberta T2N 1N4 Canada}%
\affiliation{Program in Quantum Information Science, Canadian Institute for Advanced Research, Toronto, Ontario M5G 1M1 Canada}
\homepage{http://iqst.ca/people/peoplepage.php?id=4}
 \email{sandersb@ucalgary.ca}
\date{\today}% It is always \today, today,

\begin{abstract}
We develop a concept for
a waveguide that exploits spatial control of nonlinear surface-polaritonic waves.
Our scheme includes an optical cavity with four-level $\text{N}$-type atoms in a lossless dielectric placed above a negative-index metamaterial layer.
We propose exciting a polaritonic Akhmediev breather at a certain position of the interface between the atomic medium
and the metamaterial by modifying laser-field intensities and detunings.
Furthermore, we propose generating 
position-dependent polaritonic frequency combs by engineering widths of the electromagnetically induced transparency window commensurate
with the surface-polaritonic modulation instability.
Therefore, this waveguide acts as a high-speed polaritonic modulator and position-dependent frequency-comb generator,
which can be applied
to compact photonic chips. 
\end{abstract}
\pacs{Valid PACS appear here}% PACS, the Physics and Astronomy
\maketitle
Nonlinear plasmonics (and polaritonics)~\cite{kauranen2012nonlinear} in waveguide geometries are of strong interest for schemes enabling strong cross-phase modulation~\cite{PhysRevA.81.033839},
amplification and lasing~\cite{berini2012surface}, modulators~\cite{schuller2010plasmonics} and detection~\cite{brongersma2015plasmon}.
 Controlling and exciting nonlinear surface polaritons (SPs) is challenging as the strength of the nonlinear processes and their efficiency depend strongly on (metallic) nanostructure roughness~\cite{nahata2003enhanced,feth2008second} which is experimentally challenging to minimize. We circumvent
this problem by formulating an approach that spatially controls nonlinear SP waves, and we explore its application for modulation~\cite{schuller2010plasmonics} and frequency-comb generators~\cite{geng2016frequency}.

\begin{figure}
\includegraphics[width=1\columnwidth]{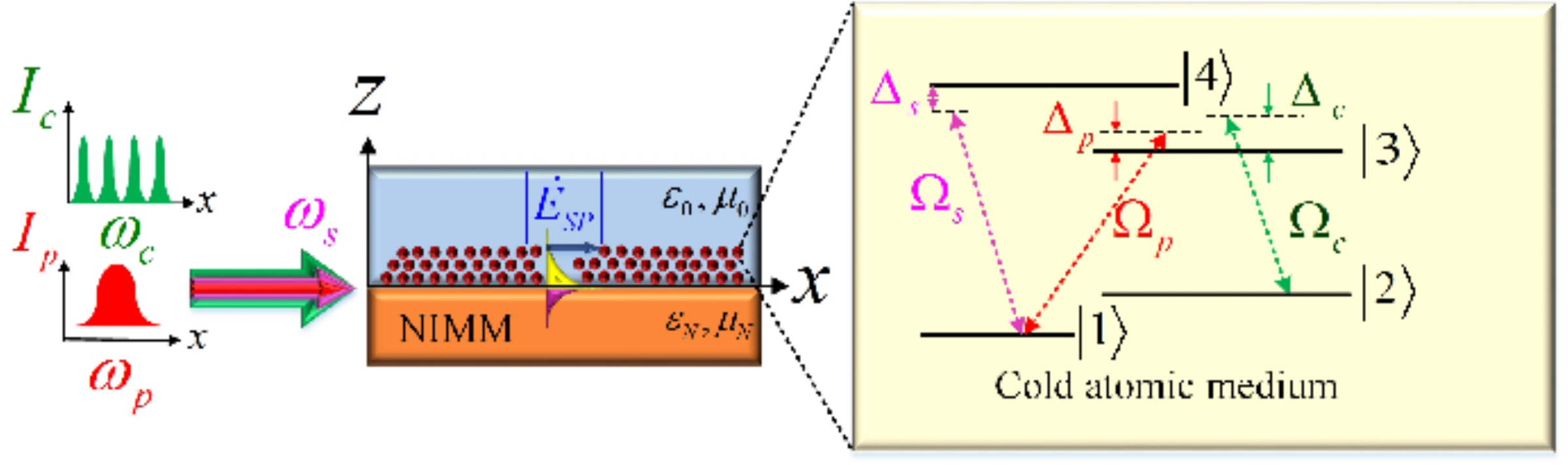}
\caption{Our proposed polaritonic waveguide comprises a 4NA medium as a thin doped layer in a lossless dielectric placed above a NIMM half-space. The waveguide is assumed placed in a cavity (not shown) and the 4NA medium driven by three co-propagating signals,  a pump signal (s), a weak probe signal (p), and a  standing wave coupling signal (c), all  assumed injected from laser beams using the end-fire coupling technique. The cavity produces a resonant mode at the coupling frequency only, so this wave is illustrated as a standing wave.}
\label{fig:Main}
\end{figure}

For spatial control of nonlinear surface-polaritonic waves,
we suggest driving four-level N-type atoms
(4NAs)~\cite{sheng2011all}
on the surface
of a negative-index metamaterial (NIMM)~\cite{xiao2010loss} as depicted in Fig.~\ref{fig:Main}.
These components are contained in a stable cavity
and serve as a nonlinear planar waveguide.
The atoms are dopants in a transparent medium
over a thickness of several dipole-transition wavelengths.
These atoms are driven by three co-propagating fields a pump signal (s), a weak probe signal (p), and a  standing wave coupling signal (c), all  assumed injected from laser beams using the end-fire coupling technique~\cite{stegeman1983excitation}.

Various approaches can be used to describe the NIMM~\cite{shalaev2007optical}.
We employ a macroscopic description
involving macroscopic permittivity and permeability,
which are inserted into the Drude-Lorentz model~\cite{PhysRevLett.101.263601,xiao2009yellow}

The 4NA is appealing
because of its giant Kerr nonlinearity
and
controllable dispersion~\cite{PhysRevA.84.053820}.
We assume that the signal~(s), probe~(p) and couple~(c) laser fields drive the $\ket4\leftrightarrow\ket1$, $\ket3\leftrightarrow\ket1$ and $\ket3\leftrightarrow\ket2$ atomic transitions, respectively.
The 4NA medium in our waveguide is assumed as $\text{Pr}^{3+}$ in $\text{Y}_2\text{Si}\text{O}_{5}$ with corresponding energy levels
\begin{align}
\ket1=&\ket{^{3}\text{H}_{4},\pm5/2},\quad \ket2=\ket{^{3}\text{H}_{4},\pm3/2},\nonumber\\
\ket3=&\ket{^{1}\text{D}_2,\pm3/2},\quad
\ket4=\ket{^{1}\text{D}_2,\pm5/2}.
\label{eq:refname1}
\end{align}
We assume inhomogeneous broadening of the atomic transitions to be in Lorentzian line shape~\cite{PhysRevA.66.063802}.
The 4NA medium has atomic density $N_\text{a}$,
natural decay rates~$\Gamma_{mn}$
and dephasing rates~$\gamma_{nm}^{\text{dep}}$
between levels~$\ket n$ and~$\ket m$~\cite{boyd2003nonlinear}. 

The signal~(s), probe~(p) and couple~(c) laser fields interact with the 4NAs in the waveguide within an optical cavity
of length~$\ell$. The signal detuning frequencies are $\Delta_\text{s,p,c}$,
and the Rabi frequencies are $\Omega_\text{s,p,c}$
with
\begin{equation}
    \Omega_\text{c}(x)
        =\Omega_\text{c}^{(0)}\sin\frac{x}{\ell}
        \label{eq:refname2}
\end{equation}
for constant Rabi-frequency coefficient~$\Omega_\text{c}^{(0)}$
and longitudinal coordinate, or position, $x$.
The fields are evanescently confined 
to the NIMM-4NA interface
with decay functions~$\zeta_\text{c,p,s}(z)$.
The decay functions are maximum at the interface, and we assume that
$\zeta_\text{c}\equiv\zeta_\text{s}\approx\zeta_\text{p}$~\cite{PhysRevA.91.023803}. 

We show that these laser driving fields 
would excite nonlinear SP waves
including Akhmediev breathers,
which is a solitary localized nonlinear wave with a periodically oscillating amplitude~\cite{akhmediev1986modulation}, and a frequency comb, as a nonlinear wave that appears briefly at specific positions.
We propose generating these nonlinear waves
by coupling the probe laser
to the dipole moment of the 4NA $\ket3\leftrightarrow\ket1$ transition,
and stability is achieved by imposing an SP low-loss condition and modifying the nonlinearity and dispersion of SPs at the interface.

Our quantitative description of the system is obtained
by solving Maxwell-Bloch equations~\cite{PhysRevA.73.020302}
based on a perturbative, asymptotic, multi-scale position~($x$)
and time~($t$)
expansion~\cite{PhysRevA.98.013825}
\begin{equation}
\label{eq:xltl}
    x_l=\varepsilon^lx,\;
    t_l=\varepsilon^lt,\;
    \varepsilon
        :=\max\left\{\left|\frac{\Omega_\text{p}}{\Omega_\text{c}}\right|,
            \left|\frac{\Omega_\text{p}}{\Omega_\text{s}}\right|\right\},
\end{equation}
for~$\varepsilon$ the perturbation scale coefficient.
Our third-order truncated solution 
yields a nonlinear Schr\"odinger equation (NLSE).
We solve and plot the Rabi frequency for the resultant surface-polaritonic Akhmediev breather
and explore Rabi-frequency dependence
as a function of various control parameters 
to identify conditions for efficient frequency-comb generation.
\begin{figure}
\includegraphics[width=1\columnwidth]{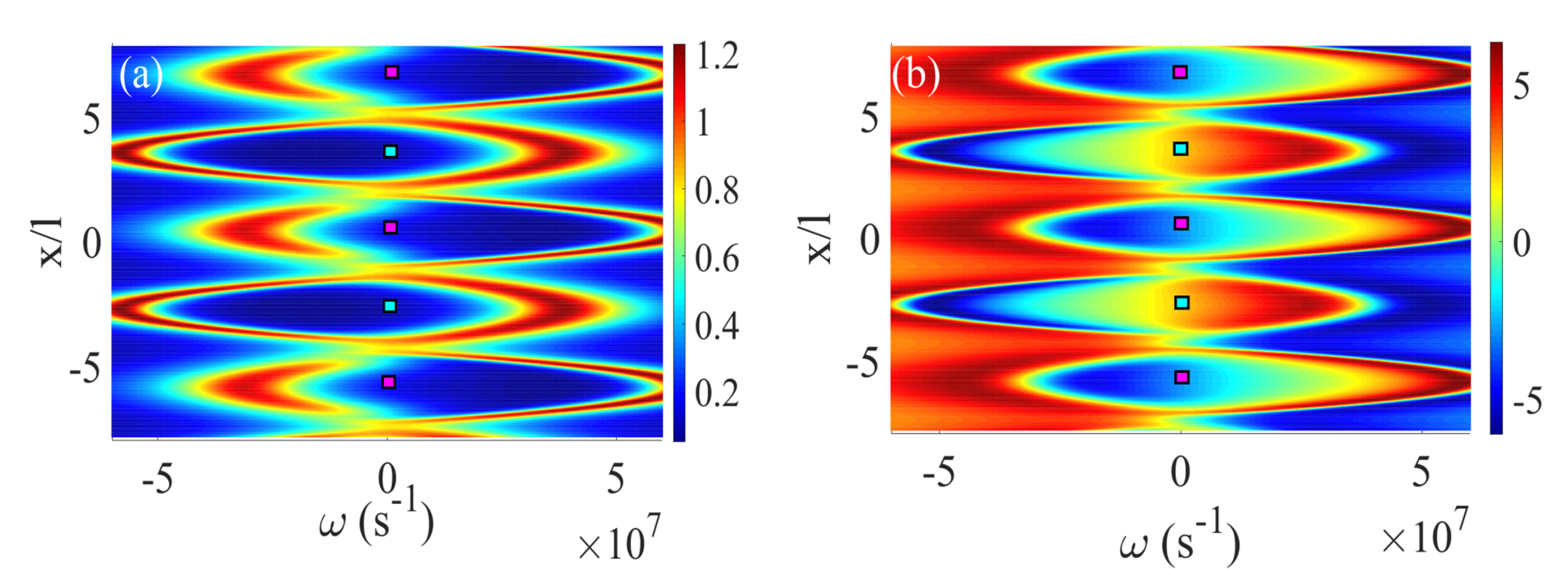}
\caption{(a)~Asymmetric absorption and (b)~dispersion spectra of the linear SPs.
Magenta dots~($x_j^{(\text{a})}$)
represent the position
for which time-periodic nonlinear waves with maximum amplitude are generated.
Blue dots~$x_j^{(\text{f})}$ show positions 
for efficient polaritonic frequency-comb generation.}
\label{fig:two}
\end{figure}

We use only~$x_{0,1,2}$ and~$t_{0,1}$
in our analysis
by ignoring three effects,
namely,
(i)~the second-order $x_1$-derivative 
due to a slowly varying amplitude~\cite{davoyan2009self},
(ii)~higher-order time scales~$t_{l>1}$
and position $x_{l>2}$,
in deriving Eq.~(\ref{eq:dimlessform}) 
by ignoring higher-order dispersion
and
(iii)~group-velocity dispersion (GVD) in the NIMM layer,
which is $10^{-5}$ times the 4NA GVD.

We treat SPs as plane waves with GVD
$K_2(\omega,x):=\partial^2 K(\omega,x)/\partial \omega^2$
for~$K(\omega,x)$
the linear dispersion.
The SP absorption coefficient is $\bar\alpha
    :=\varepsilon^2\text{Im}[K(\omega,x)]$,
and we neglect GVD
and self-phase modulation (SPM)~($W$) variability at different orders of position scales:
${\mathcal J}(x_l,\omega)$ is constant for all~$l$ 
for ${\mathcal J}\in\{K,K_2,W\}$.
Nonlinear SPs have large initial pulse width~$\tau_\text{p}$,
group velocity $v_\text{g}
=[\partial K(\omega,x)/\partial\omega]^{-1}$ and half-Rabi frequency
\begin{subequations}
\begin{align}
\label{eq:K2x}
    U_0=&\sqrt{\frac{W_\text{av}}
        {\tau_\text{p}^2K_\text{2av}}},\\
    K_{2\text{av}}=&\int_{-\ell/2}^{\ell/2}\text{d}x K_2(x),\; W_{\text{av}}=\int_{-\ell/2}^{\ell/2}\text{d}x W(x),
\end{align}
\end{subequations}
and propagate up to several nonlinear 
units of length 
given by $L_\text{N}=1/(U_{0}^{2}|W|)$ if the imaginary parts of the GVD and SPM
are much smaller than the real parts.

We replace
\begin{equation}
  t-\frac{x}{v_\text{g}}
    \mapsto
        \sigma:=\frac{\tau}{\tau_\text{p}},\;
            x\mapsto s:=\frac{x}{L_\text{N}},  
\end{equation}
 ignoring the atomic absorption due to EIT window. We normalized GVD and SPM according to $g_{\imath}(x)=\mathcal{J}(x)/\mathcal{J}_\text{av}$.
 Dynamics of the normalized SP pulse envelope~$u=[\Omega_\text{p}/U_0]\exp(-\alpha x)$ follows
\begin{align}
\label{eq:dimlessform}
    \text{i}\frac{\partial u}{\partial s}
        -\frac{g_\text{D}(x)}{2}\frac{\partial^2 u}{\partial \sigma^2}
        -g_\text{N}(x)|u|^2u\approx&0,
\end{align}
which is a dimensionless NLSE~\cite{conforti2018auto}.
 
 We propose employing the spatially modulated coupling laser for SP absorption-dispersion control during its propagation,
which we illustrate by plotting
SP absorption and dispersion in Figs.~\ref{fig:two}(a,b), respectively.
Asymmetric absorption-dispersion profiles for the position-dependent SPs are evident,
and we see the formation of multiple static EIT windows in the propagation direction by coupling laser modulation.
We reduce atomic absorption by adjusting the spatially modulated control field and other laser field intensities for the wavelength corresponding
to the $\ket3\leftrightarrow\ket1$ atomic transition (i.e., for $\omega\approx0$). Therefore,
points in the propagation direction correspond to $\omega=0$ for the multiple EIT windows seen in Figs.~\ref{fig:two}(a,b).
These EIT windows are suitable for propagating nonlinear polaritonic waves including Akhmediev breathers and frequency combs.
\begin{figure}
\includegraphics[width=1\columnwidth]{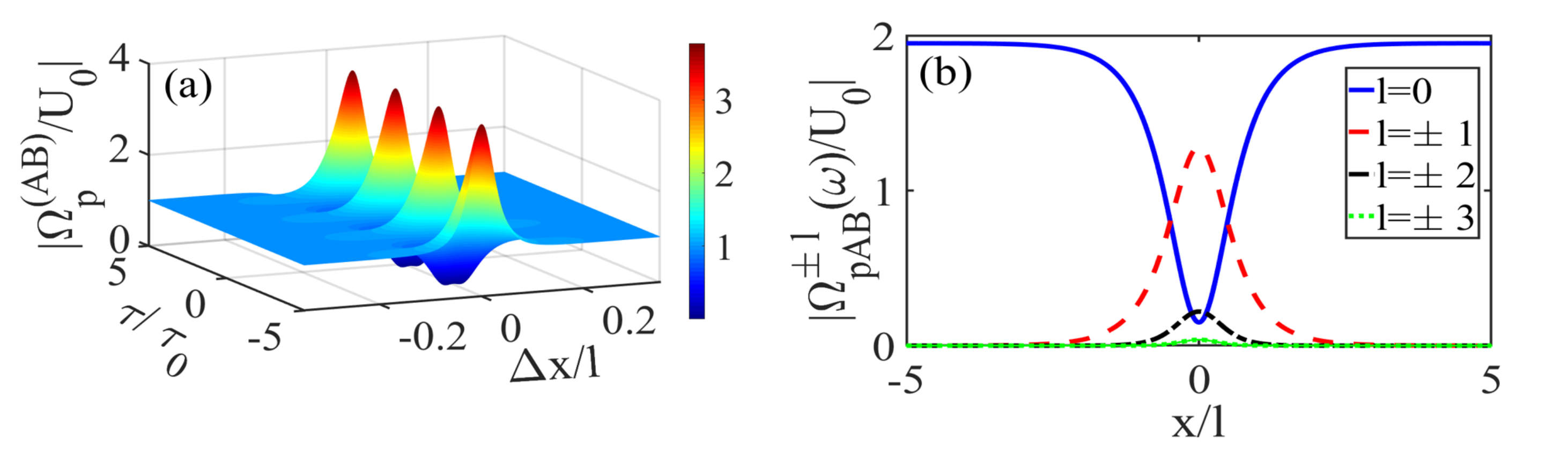}
\caption{Panel~(a) represents the GVD-SPM modulation and stable propagation of the polaritonic Akhmediev breather for $\Delta x$. Panel (b) represents the energy transfer to other polaritonic side-bands. The parameters for this simulations are: $\Omega=\sqrt{3}/2$ and $\delta\phi=0.1\pi$ and the other parameters are given in the text.}
     \label{fig:three}
 \end{figure}

 We choose realistic parameters to analyze performance of this polaritonic waveguide~\cite{wang2008all}.
 Radiative decay is quantified by $\Gamma_{31}^{\text{R}}
    =\Gamma_{43}=9~\text{kHz}$ and non-radiative decay by $\Gamma_{31}^{\text{NR}}
    \approx 6~\text{kHz}$.
 Atomic density is $N_\text{a}=4.7
    \times10^{18}~\text{cm}^{-3}$.
 We propose using a $\text{R6G}$ ring dye laser as input sources with $\lambda_l\approx 606~\text{nm}$, $\Omega_\text{s}=28~\text{MHz}$, $\Omega_\text{c}^{(0)}=80~\text{MHz}$, $\Delta_\text{s}=0.07~\text{MHz}$, $\Delta_\text{c}=0.2~\text{MHz}$,
 and $\Delta_\text{p}=0$. 
 Realistic parameters are also employed for the NIMM layer~\cite{kamli2011quantum,xiao2010loss}.

We suggest two sets of positions corresponding to $\omega\approx0$ for stable propagation of nonlinear polaritonic waves including Akhmediev breathers and frequency combs.
(i)~At positions
\begin{align}
    x_j^{(\text{a})}=(-5.68+2j\pi)\ell,\; j\in\{0,1,\dots\},
\end{align}
nonlinear SPs propagate with $v_\text{g}\approx2.91\times10^{-2}\text{c}$  within $\Delta x\approx \ell/2$,
and
\begin{subequations}
\begin{align}
K_2=&(1.45+0.09\text{i})\times10^{-15}~\text{cm}^{-1}\text{s}^2,\\ W=&(-1.47+0.11\text{i})\times10^{-15}~\text{cm}^{-1}\text{s}^2,
\label{eq:refname8}
\end{align}
\end{subequations}
respectively, are constant
with $g_\text{D}(x)\approx g_\text{N}\approx -1.01$.
Therefore, 
at these specific positions SPs propagate as polaritonic Akhmediev breathers.
(ii)~For
\begin{equation}
  x_j^{(\text{f})}
    =(-2.61+2j\pi)\ell,
\end{equation}
GVD and SPM are position-dependent within $\Delta x\approx \ell/2$
so
\begin{subequations}
\begin{align}
K_\text{2av}=&(5.87+0.25\text{i})\times10^{-18}~\text{cm}^{-1}\text{s}^2,\\ W_\text{av}=&(-1.01+0.04\text{i})\times10^{-15}~\text{cm}^{-1}\text{s}^2.
\label{eq:refname9}
\end{align}
\end{subequations}
At these points, SPs propagate with weak dispersion and strong nonlinearity as efficient polaritonic-frequency combs.
\begin{figure}
    \includegraphics[width=1\columnwidth]{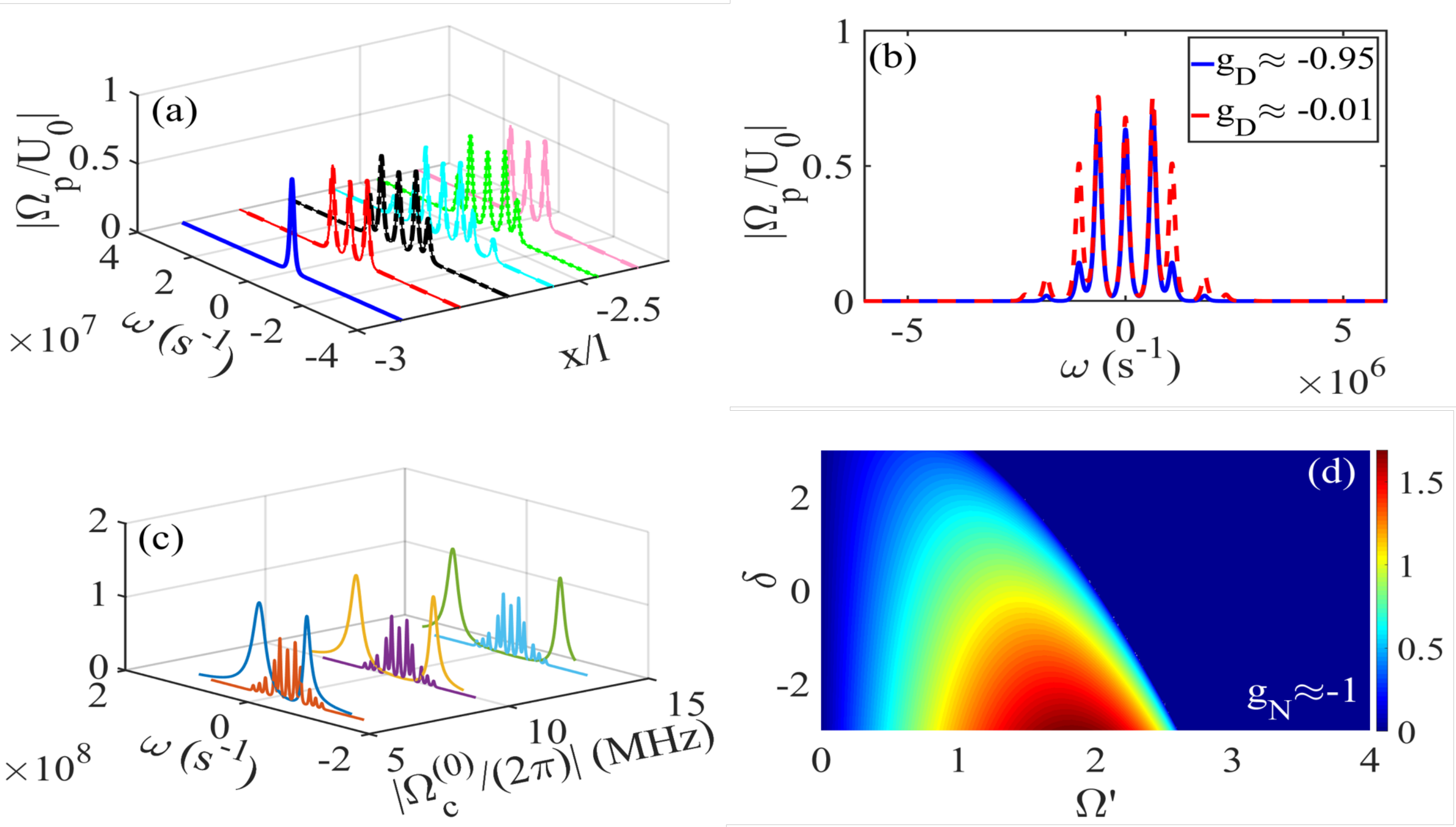}
    \caption{Panel (a) represents the propagation of the plane SP waves around the $x_0^{(\text{f})}$ dots. (b) Comparison of the Akhmediev breather $g_\text{D}=-0.95$ and efficient frequency combs $g_\text{D}\approx0.01$.  In panel (c) we apply different $\Omega_\text{c}^{(0)}$ to achieve efficient frequency combs and panel (d) represents the gain map of the polaritonic modulation instability.}
    \label{fig:four}
\end{figure}

Exploiting the correspondence between energy levels of the Bogoliubov spectrum~($E_{\pm l}$) of the uniform Bose gas with kinetic energy~($\Delta k$)~\cite{bogoliubov1947theory} and energy transferrence between nonlinear polaritonic modes in EIT windows,
we propose generating polaritonic side-bands 
with modulation frequency~$\Omega$
and growth rate~$b$ and thereby realize Akhmediev breather excitation.
Our analysis shows energy transfer from the zeroth order~($l=0$) polaritonic wave with propagation constant $b'$, $\Omega_\text{p}^0(x)=\exp(\text{i}b'\tilde{x})$,
to the first-order side-bands~($l=\pm1$) by setting $\Delta k=\Omega$ and $E_{\pm1}=b$. 

We thereby obtain the wave with amplification factor $b=-\text{i}b'$ according to
\begin{equation}
   \Omega_\text{p}^{\pm1}(x)=\text{e}^{b\tilde{x}},\;
   \tilde{x}
    :=x\left[K(\omega)+\frac{1}{2L_\text{N}}\right].
\end{equation}
Energy transmittance to third-order polaritonic side-bands~($l=\pm3$) is depicted in Fig.~\ref{fig:three}(b)
and is in accordance with energy-conservation
\begin{align}
   \left|\Omega_\text{p}^0\right|^2
+2\sum_{|l|>1}|\Omega_\text{p}^{\pm l}(x)|^2=1.
\label{eq:refname10}
\end{align}
In our scheme,
a stable polaritonic Akhmediev breather propagates for $\alpha\approx0.25$, $\Omega=0.8$ and $b=0.73$ within $\Delta x=\ell/2$ within EIT windows such that $\delta\omega_\text{EIT}\approx30~\text{MHz}$ as shown in Fig.~\ref{fig:three}(a).

Our waveguide serves as a fast-phase modulator~\cite{melikyan2014high}
according to stable polaritonic-breather propagation.
To this aim, we rewrite the surface-polaritonic Akhmediev breather solution as $\Omega_\text{p}^\text{AB}
=|\Omega_\text{p}^\text{AB}|\exp\left[\text{i}~\text{arg}\left(\Omega_\text{p}^\text{AB}\right)\right]$.
For our realistic parameters,
$\text{arg}(\Omega_\text{p}^\text{AB})\approx\pi$ which is the phase shift between initial and recovered plane SP waves after breather formation.
The time duration for the breather excitation-recurrence cycle
in our nonlinear waveguide is $\delta t=12~\text{ps}$.
Therefore, our waveguide modulates polaritonic frequencies up to a few~GHz and can be applied as a fast surface-polaritonic phase modulator. 

We propose efficient polaritonic-frequency combs by rewriting
\begin{equation}
g_\imath(x)
= g_{\imath}^\text{c}+g_{\imath}^{\text{p}}(x),\;
\imath\in\{\text{D,N}\}.
\label{eq:refname11}
\end{equation}
in terms of constant and position dependent parts.
The frequency combs can be excited at specific positions~$x_j^{(\text{f})}$,
where nonlinear SPs exhibit low GVD~($|g_\text{D}^{\text{p}}(x)|\ll1$) and strong nonlinearity~($|g_\text{N}^{\text{p}}(x)|\approx1$).
Therefore, we neglect GVD and replace $g_\text{D}[\partial^2u/\partial\sigma^2]\mapsto0$ (\ref{eq:dimlessform}) and assume $g_\text{N}^{\text{c}}\approx-1$. 
The resultant expression admits an initial SP wave with input power $P_0$ of the form
\begin{equation}
    u(x)
    =\sqrt{P}_0\exp\left[-\text{i}P_0\int\text{d}x'g_\text{N}(x')\right].
    \label{eq:refname12}
\end{equation}
We claim that stable propagation of nonlinear SPs in the weak-dispersion limit depends on the EIT-window widths and the normalized nonlinear coefficient~$g_\text{N}^{\text{p}}(x)$.

We propose efficient surface-polaritonic frequency combs by SP propagation along the interface shown in Fig.~\ref{fig:three}(a)
with $\delta\phi=0.1\pi$,
$P_0\approx10~\mu W$.
Then we numerically solve the NLSE together with initial condition~(\ref{eq:refname12}) within $-3\ell<x_0^{(\text{f})}<-2.5\ell$.
We obtain a modulated EIT window, strong nonlinearity and consequently efficient polaritonic frequency combs. Specifically, for $x_0^{(\text{f})}$ with $\delta\omega_\text{EIT}=25~\text{MHz}$ and $g_\text{D}\approx0.01$,
frequency combs up to $ \delta\omega_\text{comb}\approx 11.2~\text{MHz}$ with stability $|\Omega_\text{p}(x)|\approx0.87|\Omega_{p}(x=0)|$ are excited.
However, outside the EIT window, the generated polaritonic combs are highly unstable due to high atomic absorption.

This model allows us to develop a condition to generate efficient surface-polaritonic frequency combs via position-dependent GVD and SPM. To this aim, we consider $\Delta x=x_{*}=\varepsilon$ as a small propagation length and add a perturbative term to Eq.~(\ref{eq:refname12}) of the form
\begin{align}
    u(x,t)
    =&\left\{\sqrt{P_0}+\varepsilon p(x)\text{e}^{\text{i}\omega\tau/\tau_\text{p}}\right\}\nonumber \\ 
    &\times\exp\left[-\text{i}P_0\int_{-\frac{\ell}{2}}^{\frac{\ell}{2}}\text{d}x'g_\text{N}(x')\right].
    \label{eq:refname13}
\end{align}
We also expand the SP-wave perturbation frequency around the EIT-window centre~($\omega_{*}$) as a function of the relative polaritonic frequency comb mode number~($\nu$) in the presence of SPs dispersion
\begin{equation}
  \omega=\omega_{*}+\mathcal{K}_1(x)\nu+\frac{\mathcal{K}_2(x)}{2}\nu^2+\cdots
  \label{eq:conditionfrequencycomb}
\end{equation}
with~$\{\mathcal{K}_{l>2}\}$ related to higher-order dispersion.
Efficient frequency combs are generated by suppressing higher-order dispersion~(\ref{eq:conditionfrequencycomb}); i.e., $|\mathcal{K}_2(x)|\ll\text{c}|\mathcal{K}_1(x)|^2$.
Specifically, at $x_0^{(\text{a})}$,
$\mathcal{K}_1\approx 0.35$, $\mathcal{K}_2\approx2.16\times10^{-10}$ and $|\mathcal{K}_2/(c\mathcal{K}_1^2)|\sim10^{-18}\ll1$,
which yields efficient polaritonic-frequency combs as shown in Fig.~\ref{fig:four}(a).

We vary the coupling-laser intensity for experimental control of EIT-window widths,
leading to efficient polaritonic frequency combs
shown clearly for $g_\text{D}=0$
and $\delta\phi=0.1\pi$
with initial condition~(\ref{eq:refname13}) solving the NLSE numerically around $x_0^{(\text{f})}$.
The number of frequency combs increases by modulating coupling-laser intensity and by engineering the EIT-window widths shown in Fig.~\ref{fig:four}(c).
Comparing our frequency combs to polaritonic Akhmediev breather reveals that nonlinear waves generated at our propose position are more efficient than frequency combs excited by Akhmediev breathers, as seen 
in Fig.~\ref{fig:four}(b)).     

We describe the excitation of nonlinear surface-polaritonic waves including polaritonic Akhmediev breather and frequency combs by employing the concept of pass-band polaritonic modulation instability. We assume the initial SPs with dispersion length $L_\text{D}\approx L_\text{N}$ according to
\begin{align}
    u(x,t)
        =&u_0\text{e}^{\text{i}(k+K(\omega)+1/2L_\text{D})x
            +\text{i}\omega\tau/\tau_\text{p}}, 
\label{eq:refname15}
\end{align}
with
\begin{align}
    k=&g_{\text{av}i}|u_0|^2\text{e}^{2\text{Im}[K(\omega)]x}-\frac{\omega^2}{2}
        -K(\omega)-\frac{1}{2L_\text{N}}.
\label{eq:refname16}
\end{align}
Moreover, we assume $\Omega'$ as a modulation frequency, $\delta=(\omega_{*}-\omega)/\omega_{*}$ as the normalized perturbed frequencies and $\kappa$ as a modulation parameter in the propagation direction.
We have perturbed the SP waves in terms of $p(x),q(x)\ll1$ as
\begin{align}
u(x,t)=u_0\left[1+p(x)\text{e}^{-\text{i}\Omega'(\kappa \tilde{x}-\tau)}
        +q^{*}(x)\text{e}^{\text{i}\Omega'(\kappa^{*} \tilde{x}-\tau)}\right].
\label{eq:refname17}
\end{align}
We also expand the SPs linear dispersion and the nonlinear coefficient as a power series of the normalized perturbation frequency
\begin{equation}
  K(\delta)=K_0+K_1\delta+\frac{K_2}{2}\delta^2+\mathcal{O}(\delta^3), \; g_\text{N}\approx g_\text{0N}+g_\text{N1}\delta,
\end{equation}
and linearize the NLSE using Eq.~(\ref{eq:refname17}) in the weak perturbation limit~\cite{chen2017versatile}.
The perturbed-wave dispersion relation is  
\begin{align}
\left[\kappa+\tilde{K}_0+(K_1-1)\delta+\frac{K_2}{2}\delta^2\right]^2\nonumber\\ +(g_\text{0N}+g_\text{N1}\delta)|u_0|^2-\frac{\Omega'^2}{4}=&0,
\label{eq:refname18}
\end{align}
with $\tilde{K}_{0}=K_{0}+1/2L_\text{N}$.
The gain map for the perturbed-polaritonic waves,
shown in Fig.~\ref{fig:four}(d),
demonstrates that nonlinear-polaritonic waves are excited in EIT windows with $|\delta|<\delta_\text{EIT}$ and $0.5<\Omega'<1$ corresponding to pass-band polaritonic modulation instability. 

In summary,
we introduce a waveguide that exploits spatial control to excite nonlinear-polaritonic waves including Akhmediev breathers and frequency combs
as specific cases.
We propose a stable cavity comprising 4NAs in a lossless dielectric above the NIMM layer,
on which SPs propagate. The 4NA medium is driven by three co-propagating signals,  a pump signal (s), a weak probe signal (p), and a  standing wave coupling signal (c), all  assumed injected from laser beams using the end-fire coupling technique.
We propose stable excitation of polaritonic Akhmediev breathers and energy transfer to other polaritonic side-bands at certain position of NIMM-4NA interface
by modifying laser-field intensities and detunings through GVD-SPM modulation.
Moreover, we demonstrate efficient polaritonic frequency-comb generation at a specified position of the waveguide by engineering EIT-window widths and decreasing GVD commensurate with the pass-band regime 
for polaritonic modulation instability.
Our proposed waveguide has been analzyed for experimentally feasible conditions
and should act as a high-speed polaritonic phase modulator and efficient frequency-comb generator.
\bibliography{ref}
\end{document}